\documentclass[11pt]{article}
\usepackage{a4wide}
\usepackage{amsmath,amssymb,amsthm,amscd}
\usepackage[pdftex]{graphicx}
\pdfoutput=1

\usepackage[all]{xy}

\newcommand{\bb}[1]{\begin{equation}\label{#1}}
\newcommand{\ee}{\end{equation}}
\newcommand{\vc}[1]{{\bf #1 }}

\newcommand{\diag}{\operatorname{diag}}

\newcommand{\x}{\pmb\xi}

\def\identy{{\mathsurround0pt\mathchoice{\textidenty}{\textidenty}{\scptidenty}{\scptidenty}}}
\def\scptidenty{\setbox0\hbox{$\scriptstyle1$}\bothidenty}
\def\textidenty{\setbox0\hbox{$1$}\bothidenty}
\def\bothidenty{\rlap{\hbox to.97\wd0{\hss\vrule height.06\ht0 width.82\wd0}}
 \copy0\rlap{\kern-.36\wd0\vrule height1.05\ht0 width.05\ht0}\kern.14\wd0}

\begin{document}
\title{A Mechanistic Dynamic Emulator}

\author{C. Albert\footnote{Eawag, aquatic research, 8600 D\"ubendorf, Switzerland.}}
\maketitle

\begin{abstract}
In applied sciences, we often deal with deterministic simulation models that are too slow for simulation-intensive tasks such as calibration or real-time control.
In this paper, an emulator for a generic dynamic model, given by a system of ordinary non-linear differential equations, is developed.
The non-linear differential equations are linearized and Gaussian white noise is added to account for the non-linearities.
The resulting linear stochastic system is conditioned on a set of solutions of the non-linear equations that have been calculated prior to the emulation.
A path-integral approach is used to derive the Gaussian distribution of the emulated solution.
The solution reveals that most of the computational burden can be shifted to the conditioning phase of the emulator and the complexity of the actual emulation step  
only scales like $\mathcal O(Nn)$ in multiplications of matrices of the dimension of the state space. Here, $N$ is the number of time-points at which the solution is to be emulated and $n$ the number of solutions the emulator is conditioned on.

The applicability of the algorithm is demonstrated with the hydrological model logSPM.

\end{abstract}


\section{Introduction}

In applied sciences, we often have deterministic simulation models at hand that are, although quite accurate, too slow for many simulation-intensive tasks such as calibration or real-time control.
The purpose of an {\em emulator} (e.g. \cite{kennedy_2001_bayesiancalibration}) is a fast interpolation of the response surface of the model.
Therefore, the slow deterministic simulation model is simplified and noise is added to account for the errors due to simplification. The resulting fast stochastic model is then conditioned with outputs from the simulation model that have been produced off-line, that is, prior to the actual emulation.

In this paper, we focus on {\em dynamic} models, i.e., models described by ODE's whose outputs are given by time-series. 
Treating time as an additional output component or as an additional input \cite{conti_2010_emulationmultioutput} and applying a standard Gaussian emulator
leads to an emulation time that grows quadratically with the number of time points, which is inefficient if the number of time points is large.
Emulators for the time-stepping \cite{bhattacharya_2007_emulation}, \cite{conti_2009_emulationofdynamiccomputercodes} have the disadvantage that the whole state-space must be emulated if one wants to retain the Markov property of the process.
Simplified models in the form of stochastic linear models \cite{liu_2009_dynamicemulator} or linear combinations of (wavelet) basis functions \cite{bayarri_2007_validationwithfunctionaloutput} have been used as well.
But none of the mentioned approaches uses knowledge about the dynamics of the simulation model, which we might retrieve, e.g., by linearizing its equations.
Recently, Reichert et al. \cite{reichert_2011_dynemulator} developed a dynamic emulator whose underlying simplified model is a linear stochastic process that captures the first order dynamics of the model. That is, their emulator is to some extent {\em mechanism-based} and not merely statistical. Furthermore, they applied a Kalman filter in order for the complexity to grow only linearly with the number of time points.

The intention behind this paper is to further improve the computational efficiency of the emulator presented in \cite{reichert_2011_dynemulator}.
I start with a time-continuous linear stochastic process whose drift is assumed to be given by the linearization of the simulation model and whose noise is assumed to account for the non-linearities. This is the same as in \cite{reichert_2011_dynemulator}, except that in \cite{reichert_2011_dynemulator} the noise is not integrated between time steps.
For piece-wise constant input, I derive an analytic solution for the Gaussian distribution describing the emulated output. For this purpose, a path-integral approach seems to be adequate.
The analytic solution reveals that most of the computational burden can be shifted to the conditioning phase of the emulator so that we are left with a computational complexity for the emulation step that grows like $\mathcal O(Nn)$ in multiplications of matrices of the dimension of the state space, $m$. Here, $N$ is the number of time points and $n$ the number of simulation outputs on which the stochastic model is conditioned.
This is quite a substantial improvement of the algorithm presented in \cite{reichert_2011_dynemulator}, whose emulation step needs $\mathcal O(N)$ multiplications of matrices of dimension $nm$ as well as inversions of matrices of dimension $m$.

Just like in \cite{reichert_2011_dynemulator}, the algorithm is then tested with the hydrological model logSPM.

\section{A Generic Dynamic Emulator}
Consider a {\em state space} $V$ of dimension $m$, whose elements shall be denoted by $\x$, and a {\em deterministic simulation model}, given by a system of ordinary differential equations
\begin{equation}\label{simulation}
\dot\x(t)=f(\x(t),\vc x(t))\,,
\end{equation}
where $\vc x\in W$ denotes inputs and/or parameters and can be time-varying.
Subsequently, I refer to $\vc x$ as input and usually omit its time argument.

The idea behind the emulator is, firstly, to {\em linearize} eq. (\ref{simulation}) and pack all the non-linearities into a {\em noise} term that is modeled with a standard Wiener process $\pmb\eta(t)$ (i.e. Gaussian white noise).
The covariance of the noise, $C$, is assumed to be independent of the input. 
Thus, 
the linear stochastic approximation to (\ref{simulation}) is given by the system of linear stochastic differential equations
\begin{equation}\label{linarization}
\dot\x(t)=A(\vc x)\x(t)+\vc b(\vc x)+C{\pmb\eta}(t)\,.
\end{equation}
Secondly, $n+1$ {\em replica of the system are coupled}. 
Therefore, replace $V$ by $V\otimes \mathbb R^{n+1}$ and $W$ by $W=W\otimes \mathbb R^{n+1}$. Henceforth, vectors without indices will denote elements of these extended spaces.
The $n+1$ replica are associated with $n+1$ different inputs, $\vc x^\alpha$.
The first $n$ inputs are those for which solutions of (\ref{simulation}) are calculated that will be used for the conditioning while the $(n+1)$th input is the one for which a solution is to be emulated.
The replica are assumed to couple through the noise term only. Thus, $A(\vc x)$ and $\vc b(\vc x)$ now denote the tensors
\begin{align}
 A^{\alpha}_{\beta}({\vc x})=A({\vc x}^{\alpha})\delta^{\alpha}_{\beta}\,,\\
{\vc b}^{\alpha}({\vc x})=\vc b({\vc x}^\alpha)\,.
\end{align}
The closer the inputs (measured with some metric $\rho$ on $W$) the stronger the associated replica are assumed to couple. Hence, set
$$
\tilde C({\vc x})= C\otimes R({\vc x})\,,
$$
with 
$$
R^{\alpha\beta}({\vc x})=\exp(-\rho(\vc x^\alpha,\vc x^\beta)/2)\,.
$$
Thus, the emulator is described by the $nm$ coupled linear stochastic differential equations
\begin{equation}\label{emulator}
\dot{\x}(t)= A({\vc x})\x(t)+ {\vc b}({\vc x})+ \tilde C({\vc x}){\pmb\eta}(t)\,.
\end{equation}

Next, I derive the probability density of $\x(t)$ on the space of paths $[t_0,t_N]\longrightarrow  V\otimes \mathbb R^{n+1}$, with initial condition
\bb{bdy}
\x(t_0)=0\,.
\ee
It reads
\begin{multline}\label{pathmeasure}
P[\x(t)]\propto\exp
\left[-\frac{1}{2}\int_{t_0}^{t_N}
\left(\dot{\x}(t)- A({\vc x})\x(t)-{\vc b}({\vc x})\right)^{\dag}
(\tilde C\tilde C^T)^{-1}({\vc x})
\left(\dot{\x}(t)- A({\vc x})\x(t)-{\vc b}({\vc x})\right)
dt\right]\\
=\exp\bigg[
-\frac{1}{2}\int_{t_0}^{t_N}
\left(\x(t)-D^{-1}{\vc b}({\vc x})\right)^{\dag}
(D^{\dag}(\tilde C\tilde C^T)^{-1}D)({\vc x})
\left(\x(t)-D^{-1}{\vc b}({\vc x})\right)
dt\bigg]\,,
\end{multline}
where 
\begin{equation}\label{DDdag}
D=\frac{\partial}{\partial t}- A({\vc x})\,.
\end{equation}
To proceed, I need to determine the {\em Green's function} of $D$.
The most general solution of 
\bb{Greens1}
D(t)G(t,t')=\delta(t-t')\,
\ee
is given by (see, e.g., \cite{kleinert_2009_pathintegrals} Chapter 3.3)
\begin{equation}\label{Greens}
G(t,t')=(\bar\Theta(t-t')+c(t'))\mathcal P \exp\left[\int_{t'}^t A(\vc x(\tau))d\tau\right]\,,
\end{equation}
where $\bar\Theta(t-t')$ is the regularized Heavyside function with $\bar\Theta(0)=1/2$ and $\mathcal P$ denotes path-ordering of the exponential.
The function $c(t')$ is determined by the boundary condition (\ref{bdy}), which entails
$$
G(t_0,t)=0\,,
$$
and translates into
$$
c(t')\equiv 0\,.
$$
Then, the adjoint Green's function reads as
\bb{Gdag}
G^{\dag}(t,t')=\bar\Theta(t'-t)\mathcal P \exp\left[-\int_{t'}^{t} A^T(\vc x(\tau))d\tau\right]\,.
\ee
Now, I calculate the {\em correlation functions} for two replica at two different time points, as expressed by the $m\times m$ matrices
\bb{twopointfunction}
\tilde\Sigma_{ij}^{\alpha\beta}=\langle\x^\alpha(t_i)\otimes\x^\beta(t_j)\rangle
=
Z^{-1}\int \exp\left[
-\frac{1}{2}\x^\dag(t)D^\dag
(\tilde C\tilde C^T)^{-1}
D({\vc x})\x(t)
\right]
\x^\alpha(t_i)\otimes\x^\beta(t_j)
\mathcal D\x\,,
\ee
with
$$
Z=\int \exp\left[
-\frac{1}{2}\x^\dag(t)D^\dag(\tilde C\tilde C^T)^{-1}D({\vc x})\x(t)
\right]
\mathcal D\x\,.
$$
Using (\ref{Greens}) and (\ref{Gdag}) one finds that, for $t_i\geq t_j$,
\begin{align}\label{tildeSigma}
\tilde\Sigma_{ij}^{\alpha\beta}
=
\int_{t_0}^{t_j}
\mathcal P\exp\left[\int_{t'}^{t_i}A(\vc x^\alpha(\tau))d\tau\right]
(\tilde C\tilde C^T)^{\alpha\beta}(\vc x(t'))
\mathcal P\exp\left[-\int_{t_j}^{t'}A^T(\vc x^\beta(\tau))d\tau\right]dt'\,.
\end{align}
For $t_i<t_j$, one may use the symmetry relations
\bb{transpose}
\tilde\Sigma_{ij}^{\alpha\beta}=(\tilde\Sigma_{ji}^{\beta\alpha})^T\,.
\ee

All {\em finite dimensional marginals} of (\ref{pathmeasure}) will be Gaussians.
I consider the finite-dimensional subspace of those components of the first $n$ replica that are simulated with (\ref{simulation}) and those components of the $(n+1)$th replica that shall be emulated, both at time points $t_0<t_1<\dots<t_N$.
Therefore, I introduce the operator $H(\vc x)$ that is defined as
\begin{equation}\label{measurement}
(H(\vc x)[\x(t)])^{\alpha}_i=H(\vc x^{\alpha}(t_i))\x^{\alpha}(t_i)\,,
\end{equation}
where, on the r.h.s., $H(\vc x^{\alpha}(t_i))=:H^\alpha_i$ denotes matrices of constant rank $m'<m$.
The image of (\ref{measurement}) is supposed to be determined by eq.
\bb{y}
(H(\vc x)[\x(t)])^{\alpha}_i=\vc y_i^\alpha\,.
\ee
Integrating out all degrees of freedom that are not determined by (\ref{y}) yields the Gaussian distribution\footnote{To keep the notation simple, we will omit the dependence of $H$, $D$, $\tilde C$, and $\vc b$ on $\vc x$ subsequently.}
\begin{multline}
Z^{-1}\int P[\x(t)]\delta( H[\x(t)]-{\vc y})\mathcal D\x\\
\propto\int\exp\left[-\frac{1}{2}\int\x^{\dag}(t)D^{\dag}
(\tilde C\tilde C^T)^{-1}
D \x(t)dt\right]
\delta( H[\x(t)]-( {\vc y}- HD^{-1}{\vc b}))\mathcal D\x\\
\propto\exp\left[
-\frac{1}{2}({\vc y}- HD^{-1}{\vc b})^{\dag}
\Sigma^{-1}
({\vc y}- HD^{-1}{\vc b})\right]\,,
\end{multline}
with covariance matrix
\begin{equation}
\Sigma= H D^{-1}(\tilde C\tilde C^T)(D^{\dag})^{-1} H^{\dag}\,,
\end{equation}
and mean
\bb{mean}
\vc z= HD^{-1}{\vc b}\,.
\ee
The covariance matrix is a square matrix of dimension $(n+1)Nm'$, whose $m'\times m'$ blocks are given by the equations
\begin{equation}\label{Sigma}
\Sigma^{\alpha\beta}_{ij}
=H^{\alpha}_i
\tilde\Sigma_{ij}^{\alpha\beta}
(H^{\beta}_j)^T\,,
\end{equation}
with $\tilde\Sigma$ as defined by (\ref{tildeSigma}) and (\ref{transpose}).

Finally, I determine the {\em Gaussian distribution for the $(n+1)$th replica (the online system)}, conditioned on the simulations $\vc y_i^a$, for $i=1,\dots,N$ and $a=1,\dots,n$.
Therefore, I split the $(n+1)$th replica off, writing $\Sigma$ as the block matrix
\begin{equation}\label{block}
\Sigma=\left(
\begin{array}{c|c}
\Sigma^{n+1,n+1}	&	\Sigma^{n+1,.}\\
\hline\\
\Sigma^{.,n+1}	&	\Sigma'
\end{array}
\right)\,.
\end{equation}
Mean and covariance matrix of the online system are given by equations
\begin{align}
\bar{\vc y}&=\vc z^{n+1}
+\Sigma^{n+1,a}(\Sigma')^{-1}_{ab}
({\vc y}^b-\vc z^b)\,,\label{EmMean}\\
\bar\Sigma&=\Sigma^{n+1,n+1}-\Sigma^{n+1,a}(\Sigma')^{-1}_{ab}\Sigma^{b,n+1}\,,\label{EmCov}
\end{align}
where $a$ and $b$ run from 1 to $n$ only. 
Note that eq. (\ref{EmMean}) is translational invariant. Thus, one may replace condition (\ref{bdy}) by 
\bb{bdycond2}
\x(0)=\x_0\,.
\ee

In the remainder of this chapter, a recursive procedure of calculating (\ref{EmMean}) 
and (\ref{EmCov}) 
is developed.
Due to path-ordering eqs. (\ref{tildeSigma}) and (\ref{Sigma}) can be written as
\begin{equation}\label{Sigma1}
\Sigma_{ij}^{\alpha\beta}
=H_i^\alpha
\bigg(
\sum_{k=0}^{j-1}
h^\alpha_{i-1}\dots h^\alpha_{k+1}
g^{\alpha\beta}_{k}
h^{\dag\beta}_{k+1}\dots h^{\dag\beta}_{j-1}
\bigg)
(H^{\beta}_j)^T\,,
\end{equation}
with
\begin{align}
g^{\alpha\beta}_{k}
&=
\int_{t_k}^{t_{k+1}}
\mathcal P\exp\left[\int_{t'}^{t_{k+1}}A({\vc x}^\alpha(\tau))d\tau\right]
(\tilde C\tilde C^T)^{\alpha\beta}(t')
\mathcal P\exp\left[-\int_{t_{k+1}}^{t'}A^T({\vc x}^\beta(\tau))d\tau\right]\,,\label{g2}\\
h^\alpha_l
&=\mathcal P\exp\left[\int_{t_l}^{t_{l+1}}A({\vc x}^\alpha(\tau))d\tau\right]\,,\label{h2}\\
h^{\dag\alpha}_l&=\mathcal P\exp\left[-\int_{t_{l+1}}^{t_l}A^T({\vc x}^\alpha(\tau))d\tau\right]\,.
\end{align}
The boundary conditions (\ref{bdy}) or (\ref{bdycond2}) imply the initial variances
\begin{equation}\label{ini}
\Sigma_{00}^{\alpha\beta}=0\,.
\end{equation}
In the {\em conditioning step} of the algorithm one calculates $(\Sigma')^{-1}$ and $\vc z^a$, for $a=1,\dots,n$.
For the former, use (\ref{transpose}), (\ref{Sigma}) and (\ref{ini}) and, for $j\leq i$, the recursion relations
\begin{align}
\Sigma_{i+1,j}^{\alpha\beta}
&=h^\alpha_i\Sigma_{ij}^{\alpha\beta}\label{rec}\\
\Sigma_{ii}^{\alpha\beta}
&=\Sigma_{i,i-1}^{\alpha\beta}h^{\dag\beta}_{i-1}
+g^{\alpha\beta}_{i-1}\,.\label{reca}
\end{align}
For the latter, set
$$
{\vc z}^\alpha_i=H^\alpha_i{\tilde{\vc z}}^\alpha_i\,,
$$
and use the recursion relations
\begin{equation}\label{zrec}
\tilde{\vc z}_{i+1}^{\alpha}=h_i^{\alpha}\tilde{\vc z}_i^{\alpha}
+\vc k^{\alpha}_i\,,\quad
\tilde{\vc z}_0^{\alpha}=0\,,
\end{equation}
with 
\begin{equation}\label{k2}
\vc k^\alpha_i=\int_{t_i}^{t_{i+1}}
\mathcal P\exp\left[\int_{t'}^{t_{i+1}}A({\vc x}^\alpha(\tau))d\tau\right]
{\vc b}^\alpha(t')dt'\,.
\end{equation}
Once $(\Sigma')^{-1}$ and all the $\vc z^a$ are calculated, pre-calculate, for the emulation step, the covectors
\bb{ztildeprime}
{\vc z}'{}_{ia}
:=T_{ij}^a(H_j^a)^T((\Sigma ')^{-1})^{jk}_{ab}(\vc y_k^b-\vc z_k^b)\,,
\ee
where, on the r.h.s., the $i$'s and the $a$'s are not summed over and with
\bb{T}
T_{ij}^a:=\left\lbrace
\begin{array}{cc}
h^{\dag a}_{i+1}\dots h^{\dag a}_{j-1}\,,&\quad j\geq i+2\,,\\
\identy\,,&\quad j=i+1\,,\\
0\,,&\quad\text{else}\,.
\end{array}
\right.
\ee
In the actual {\em emulation step} calculate (\ref{EmMean}) setting
\bb{ytilde}
\bar{\vc y}_i=H^{n+1}_i\tilde{\vc y}_i\,,
\ee
and using the recursion relation
\bb{rec3}
\tilde{\vc y}_{i+1}=h^{n+1}_i\tilde{\vc y}_i
+\vc k^{n+1}_i
+g^{n+1,a}_i{\vc z}'{}_{ia}\,,
\ee
with $\vc z'_{ia}$ as defined in (\ref{ztildeprime}).
In order to get the start value, $\tilde{\vc y}_1$, one needs to calculate $\Sigma^{n+1,a}_{1j}$ using (\ref{transpose}), (\ref{ini}) and the recursion relations (\ref{rec}) and (\ref{reca}).
The computational complexity of the emulation step is of the order $\mathcal O(Nn)$ in matrix multiplications of dimension $m$.
If one is interested in the variances, i.e., the diagonal elements of $\bar\Sigma$, one may derive a similar recursion formula for them.

Since path-ordered exponentials can, in general, not be calculated analytically, I consider the special case of {\em piece-wise constant input}
$$
{\vc x}^\alpha(t)={\vc x}^\alpha_i\,,\quad t_i\leq t\leq t_{i+1}\,.
$$
Then, (\ref{g2}), (\ref{h2}) and (\ref{k2}) reduce to
\begin{align}
g^{\alpha\beta}_{k}	
&=(R^{\alpha\beta}_k)^2\int_{t_k}^{t_{k+1}}e^{(t_{k+1}-t')A({\vc x}^\alpha_k)} CC^Te^{(t_{k+1}-t')(A({\vc x}^\beta_k))^T}dt'\,,\label{g}\\
h^\alpha_{l}
&=e^{(t_{l+1}-t_l)A({\vc x}^\alpha_l)}\,,\label{h}\\
\vc k^\alpha_i&=\int_{t_i}^{t_{i+1}}e^{(t_{i+1}-t')A({\vc x}^\alpha_i)}
{\vc b}_i^\alpha dt'\,.\label{k}
\end{align}
If $A(\vc x)$ is {\em diagonalizable}, functions (\ref{g}) through (\ref{k}) can be obtained analytically.
For
$$
A({\vc x}^\alpha_k)=M^\alpha_k\diag_o\left[\lambda_o^\alpha\right]
(M^\alpha_k)^{-1}\,,
$$
one gets
\bb{g1}
g^{\alpha\beta}_{k}=
(R^{\alpha\beta}_k)^2
M^\alpha_k
B^{\alpha\beta}_{k}
(M^\beta_k)^{T}\,,
\ee
with
\begin{equation}\label{B}
(B^{\alpha\beta}_{k})^p{}_q
=
\frac{\exp((t_{k+1}-t_k)(\lambda^\alpha_{k,p}+\lambda_{k,q}^\beta))-1}
{\lambda^\alpha_{k,p}+\lambda^\beta_{k,q}}
((M^\alpha_k)^{-1} C C^T((M^\beta_k)^{-1})^T)^p{}_q\,,
\end{equation}
and
\bb{h1}
h^\alpha_l=
M^\alpha_l
\diag_o\left[
		\exp((t_{l+1}-t_l)\lambda_{l,o}^\alpha)
	\right]
(M^\alpha_l)^{-1}\,,
\ee
and
\bb{k1}
\vc k^\alpha_i=
M^\alpha_i
\diag_o\left[
	\frac
	{\exp((t_{i+1}-t_i)\lambda^\alpha_{i,o})-1}
	{\lambda_{i,o}^\alpha}
\right]
(M^\alpha_i)^{-1}{\vc b}^\alpha_i\,.
\ee

If $\vc x$ is time-independent (e.g. parameters of the model) and $A(\vc x)$ diagonalizable, (\ref{Sigma}) can be calculated explicitly.
If
$$
A(\vc x^\alpha)=M^\alpha\diag_o\left[\lambda_o^\alpha\right](M^\alpha)^{-1}\,,
$$
one derives from (\ref{Sigma}) that, for $t_i>t_j$,
$$
\tilde\Sigma_{ij}^{\alpha\beta}
=(R^{\alpha\beta})^2M^\alpha B^{\alpha\beta}_{ij}(M^\beta)^T\,,
$$
where
\begin{multline}\label{Sigmaconst}
(B^{\alpha\beta}_{ij})^p{}_q
=
((M^\alpha)^{-1} C C^T((M^{\beta})^T)^{-1})^p{}_q
\int_{t_0}^{t_j}\exp[t_i\lambda^\alpha_p+t_j\lambda^\beta_q-t'(\lambda^\alpha_p+\lambda^\beta_q)]dt'\\
=
((M^\alpha)^{-1} C C^T((M^{\beta})^T)^{-1})^p{}_q
\frac
{\exp\left((t_i-t_0)\lambda^\alpha_p+(t_j-t_0)\lambda^\beta_q\right)
-\exp\left((t_i-t_j)\lambda^\alpha_p\right)}
{\lambda^\alpha_p+\lambda^\beta_q}\,.
\end{multline}

\section{Hydrological Application}
\begin{figure}[b]
\centering
\includegraphics[width=70mm]{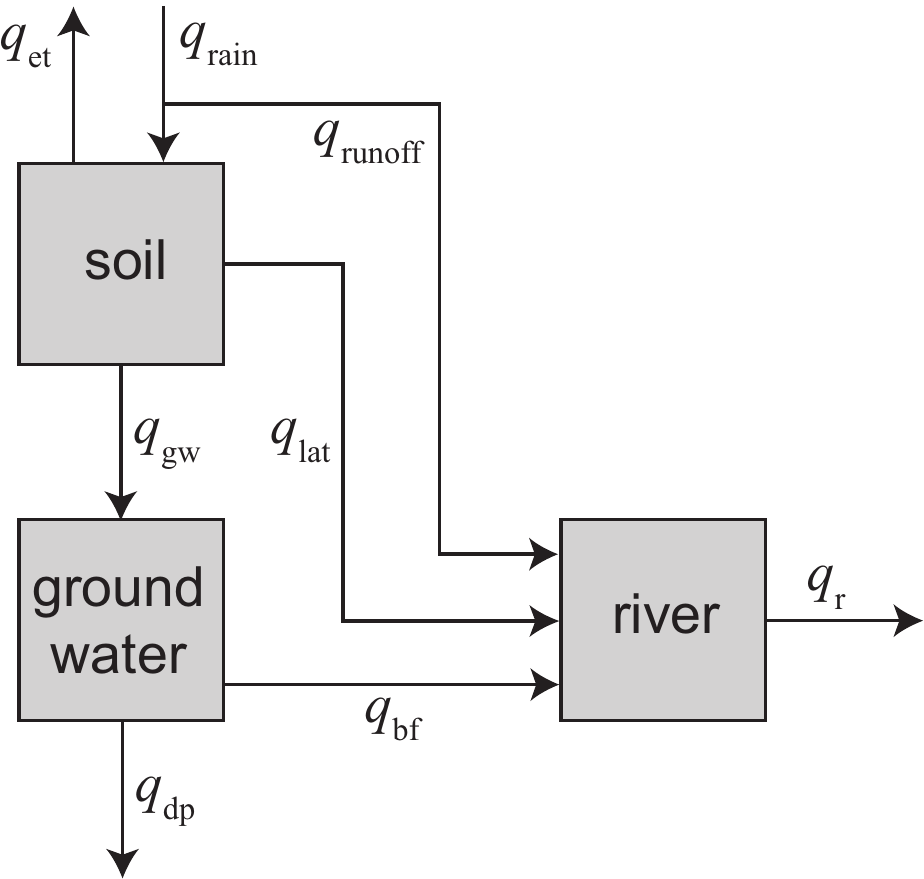}
\caption{Visualization of the fluxes in the model logSPM.
Taken from J. Comp. Stat. and Data Analysis.}\label{Fig1}
\end{figure}

In this section, the algorithm developed in the last section is tested with a simple hydrological model called logSPM \cite{kuczera_2006_bayesiantotalerroranalysishydrology}.
The state vector of this model is three-dimensional,
\begin{equation}
{\pmb\xi}=(h_s,h_{gw},h_r)^T\,,
\end{equation}
and describes the amount of water stored in the soil, the ground-water and the river.
The dynamics is described by the system of ordinary differential equations
\begin{align}\label{SPMmodel}
\dot h_s&=q_{rain}-q_{runoff}-q_{et}-q_{lat}-q_{gw}\,,\\
\dot h_{gw}&=q_{gw}-q_{bf}-q_{dp}\,,\\
\dot h_r&=q_{runoff}+q_{lat}+q_{bf}-q_r\,,
\end{align}
and visualized in Fig. \ref{Fig1}.
The fluxes are given by the equations
\begin{alignat}{2}\label{SPMfluxes}
q_{rain}&=i_{rain}(t)\,,&\quad
q_{runoff}&=f_{sat}i_{rain}(t)\,,\\
q_{et}&=f_{et}i_{pet}(t)\,,&\quad
q_{lat}&=f_{sat}q_{lat,max}\,,\\
q_{gw}&=f_{sat}q_{gw,max}\,,&\quad
q_{bf}&=k_{bf}h_{gw}\,,\\
q_{dp}&=k_{dp}h_{gw}\,,&\quad
q_r&=k_rh_r\,,
\end{alignat}
with the fraction of saturated area, $f_{sat}$, given by equation
\bb{fsat}
f_{sat}=\frac{1}{1+s_Fe^{-k_sh_s}}
-\frac{1}{1+s_F}\,,
\ee
and the fraction of actual evapotranspiration, $f_{et}$, given by equation
\bb{fet}
f_{et}=1-e^{-k_{et}h_s}\,.
\ee
The output of the model is the river flow, $Q_r$, given as
$$
Q_r=A_Wq_r\,,
$$
where $A_W$ is the area of watershed.

\begin{figure}[b]
\centering
\includegraphics[width=150mm]{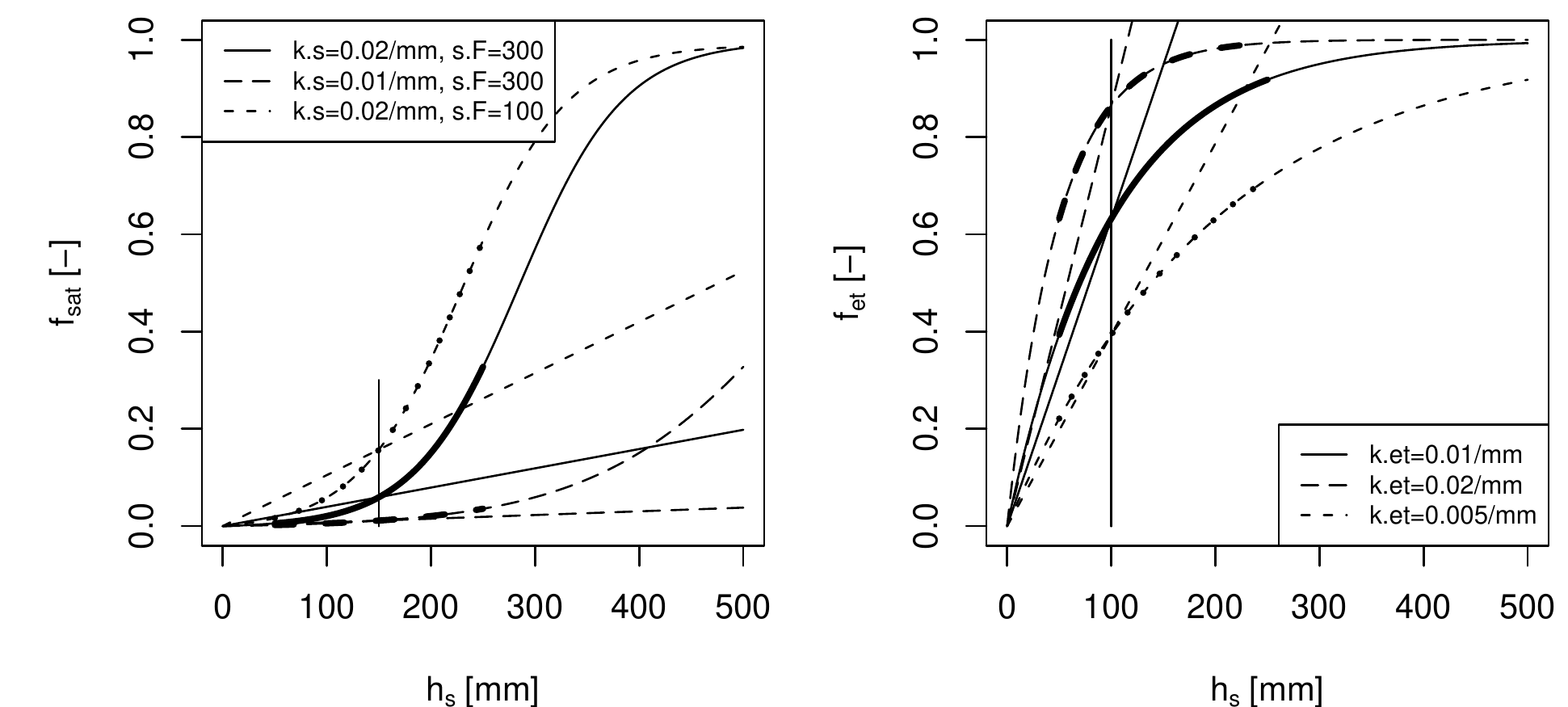}
\caption{Shape of the nonlinear functions used for describing some fluxes in the hydrological model. Fraction of saturated are, left, and fraction of actual evapotranspiration, right. The bold parts of the curves represent the range of values covered in the base simulation. The straight lines represent linearizations that intersect the nonlinear function at given values of $h_s$.
Taken from J. Comp. Stat. and Data Analysis.}\label{Fig2}
\end{figure}

The linearization of the model equations reads:
\begin{equation}
\dot{\pmb\xi}(t)=A(\vc x){\pmb\xi}(t)+\vc b(\vc x)\,,
\end{equation}
with
\begin{equation}
A(\vc x)=
\begin{pmatrix}
\lambda_1(t)&0&0\\
a&\lambda_2&0\\
c(t)&b&\lambda_3
\end{pmatrix}
\,,\quad
\vc b(\vc x)=
\begin{pmatrix}
i_{rain}(t)\\
0\\
0
\end{pmatrix}\,,
\end{equation}
and
$$
\vc x(t)=(\lambda_1(t),\lambda_2,\lambda_3,a,b,c(t),i_{rain}(t))^T\,,
$$
with
\begin{equation}
a=a_{sat}q_{gw,max}\,,\quad
b=k_{bf}\,,\quad
c(t)=a_{sat}(i_{rain}(t)+q_{lat,max})\,,
\end{equation}
and
\begin{equation}
\lambda_1(t)=-a_{sat}(i_{rain}(t)+q_{lat,max}+q_{gw,max})-a_{et}i_{pet}(t)\,,\quad
\lambda_2=-k_{bf}-k_{dp}\,,\quad
\lambda_3=-k_r\,.
\end{equation}
The functions (\ref{fsat}) and (\ref{fet}) were approximated by linear functions that intersect the nonlinear functions at $h_{s,1}$ and $h_{s,2}$, respectively. 
See Fig. \ref{Fig2}.
Therefore,
$$
a_{sat}=\frac{1}{h_{s,1}}\left(
\frac{1}{1+s_Fe^{-k_sgh_{s,1}}}
-
\frac{1}{1+s_F}
\right)\,,
$$
and
$$
a_{et}=\frac{1}{h_{s,2}}\left(
1-e^{-k_{et}h_{s,2}}
\right)\,.
$$
Only the inputs $i_{rain}$ and $i_{pet}$ are time-dependent, and, therefore, $c$, $\vc b$ and $\lambda_1$.
The observation matrices read as 
\begin{equation}
H^\alpha=
\begin{pmatrix}
0&0&0\\
0&0&0\\
0&0&-A_W\lambda_3^\alpha
\end{pmatrix}\,.
\end{equation}
I choose the Euclidean metric on the space of parameters 
$$
\mathbb R^8\ni {\pmb\theta}=(k_s,s_F,k_{et},q_{lat,max},q_{gw_max},k_{bf},k_{dp},k_r)^T\,,
$$ 
where each component is normalized with a reasonable range.
The noise $C$ was chosen to be diagonal and the diagonal entries to be a certain fraction of the initial condition $\x_0$.

\begin{figure}[b]
\centering
\includegraphics[width=150mm]{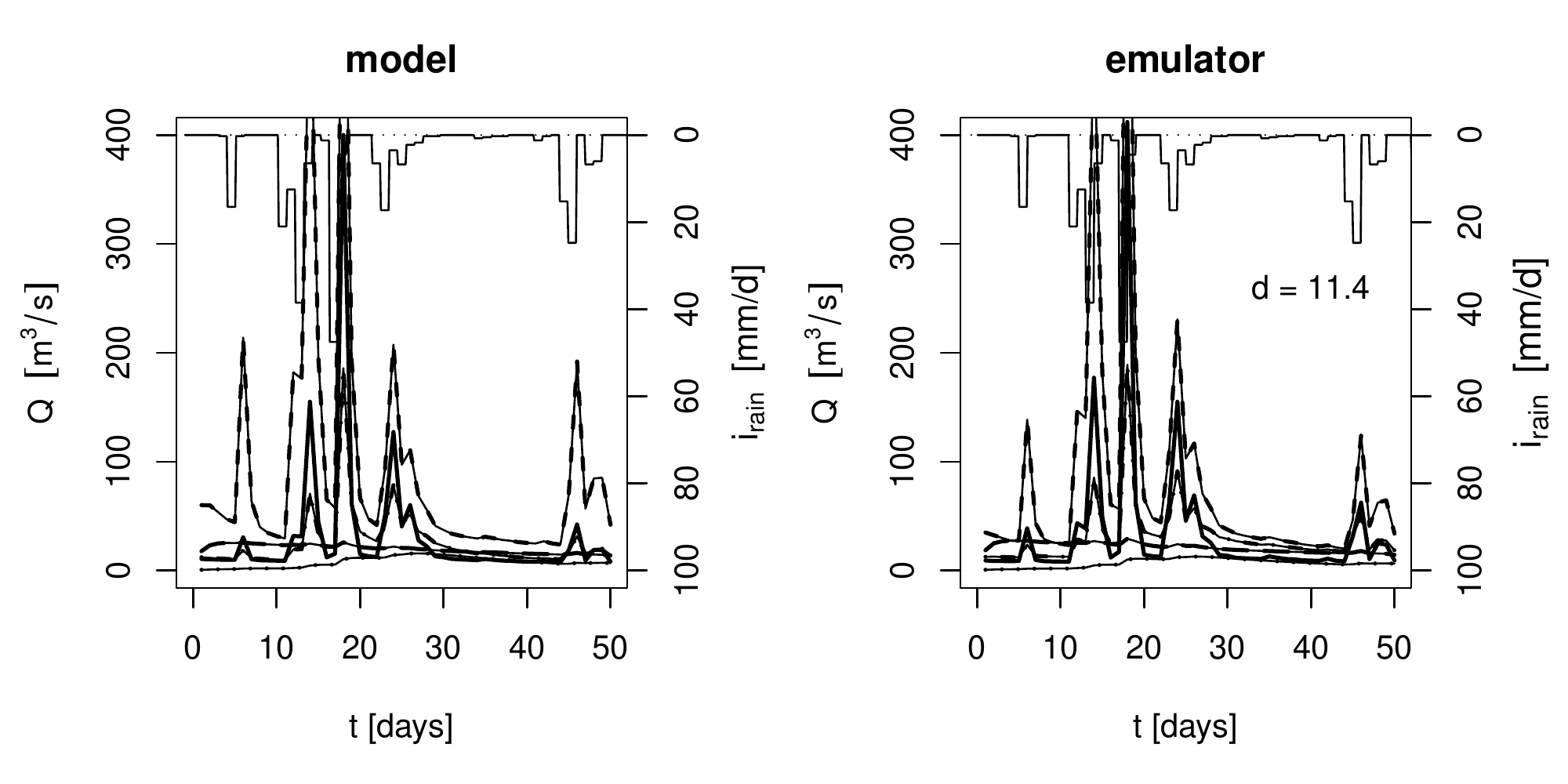}
\caption{Comparison of simulations of the full model (left) with emulations (right) for 5 randomly chosen sets of parameters. The dominant model input, rain intensity, is plotted from the top (right scale).
The emulator was conditioned with 50 sets of parameters.
The $d$-value is the square root of the mean sum of squares.
}\label{Fig3}
\end{figure}

Obviously, $A(\vc x)$ is diagonalizable:
\begin{equation}
M^{-1}(t)A(\vc x)M(t)=\diag_o[\lambda_o]\,,
\end{equation}
with
\begin{equation}
M(t)=
\begin{pmatrix}
1&0&0\\
\frac{a}{\lambda_1-\lambda_2}&1&0\\
\frac{c(\lambda_1-\lambda_2)+ab}{(\lambda_1-\lambda_2)(\lambda_1-\lambda_3)}&
\frac{b}{\lambda_2-\lambda_3}&1
\end{pmatrix}\,,
\end{equation}
and the matrices (\ref{g1}), (\ref{h1}) and (\ref{k1}) can be calculated analytically.
Plot \ref{Fig3} compares solutions of the full model with emulated solutions for 5 randomly chosen sets of parameters (that were not used for the conditioning of the emulator).
The results are very similar to those obtained in \cite{reichert_2011_dynemulator}. For an extended statistical analysis of the performance of the emulator, I refer to \cite{reichert_2011_dynemulator}. 


\section{Conclusions}

I have presented an explicit solution for the emulation of the time-series of a dynamic model.
In general, the path-ordered exponentials the solution is expressed with cannot be calculated analytically. Therefore, I resort to piece-wise constant inputs. Then, the emulator presented in this paper is the same as the one presented in \cite{reichert_2011_dynemulator}, except that I integrate the noise between time steps. For piece-wise constant input, this can be done at negligible additional cost and potentially increases the quality of the emulation.

The exact solution presented in eqs. (\ref{EmMean}) and (\ref{EmCov}) allows for an efficient numerical implementation that is of the order $\mathcal O(nN)$ in matrix multiplications of dimension $m$.
The Kalman filtering and smoothing algorithm used in \cite{reichert_2011_dynemulator} needs $\mathcal O(N)$ matrix multiplications of dimension $nm$ and matrix inversions of dimension $m$.

The disadvantage of my method, however, as compared to the one presented in \cite{reichert_2011_dynemulator} is the fact that a huge matrix of dimension $Nnm'$ needs to be inverted for the conditioning, which might be challenging both for the memory and the CPU. 

\subsection*{Acknowledgments:}
I'm indebted to Peter Reichert for many fruitful discussions about emulators as well as many lines of R Code for the presentation of the results.

\bibliographystyle{plain}
\bibliography{refs}

\end{document}